\documentclass[10pt, conference, a4paper]{IEEEtran}
\usepackage{algorithmic}
\usepackage{textcomp}
\usepackage{multirow}
\usepackage{eurosym}
\usepackage{color, soul}
\usepackage{graphicx,epsfig,epstopdf,amsmath,amssymb, amsfonts,enumerate}
\usepackage{geometry}

\begin{document}

\newgeometry{left=1.39cm, right=1.39cm, top=1.9cm, bottom=4.29cm}

\title{Spatio-Temporal Analysis of Spectrum Usage in Drone-based 5G eMBB Systems for Emergency Services} 

\author{
\IEEEauthorblockN{Galini Tsoukaneri}
\IEEEauthorblockA{\textit{Samsung Research  United Kingdom}\\
Staines-Upon-Thames - UK\\
g.tsoukaneri@samsung.com}
\and
\IEEEauthorblockN{Mythri Hunukumbure}
\IEEEauthorblockA{\textit{Samsung Research United Kingdom}\\
Staines-Upon-Thames - UK\\
mythri.h@samsung.com}
}

\maketitle

\begin{abstract}
The expectation for increased capabilities of 5G networks opens the way for new verticals to be developed, with a growing interest in providing 5G connectivity for emergency service support. 
In this paper we propose a drone-based deployment scenario, where multiple drones make up a wireless link to a 5G ground base station (BS), to provide connectivity to areas affected by emergency events. Assuming that a part of the BS's available bandwidth will be allocated to the emergency link in order to provide a high-speed and reliable connection, we present a spatio-temporal analysis of the effect of such a service on commercial 5G traffic, based on the locality, time of occurrence and severity of the event. Our work is based on recent advances in spatial analysis of BS locations, and extended datasets of emergency events in the area of greater London.

\end{abstract}


\section{Introduction}
\label{sec:intro}

The advancements of cellular technology along with the large deployments and extended coverage have paved the way for the development and deployment of new applications and services. Emergency communications are one of those applications, with network operators seeking to replace the existing stand-alone deployments (e.g. TETRA, Project $25$) with new ones over cellular networks.
A prime example in the UK is the move of the existing TETRA emergency communication system~\cite{tetra} towards an LTE-based system by $2024$.

As 5G promises to provide significantly increased data rates compared to 4G, such as those provided by \textit{enhanced Mobile Broadband (eMBB)}, there is a growing interest in developing innovative services that employ 5G deployments to provide high-speed and reliable connectivity to emergency services (e.g. police, fire brigade) on an \textit{on-demand} basis. 
Furthermore, network slicing~\cite{netSlice}, which will be natively supported in 5G, will allow for dedicated communication links to the emergency services with reliability and quality guarantees.

Inspired by the interest in supporting emergency services with 5G, in this paper we propose a drone-based service to provide temporary connectivity to services attending an emergency event, where multiple drones are equipped with 5G communication hardware, in order to form a link between a base station in close proximity and the locality of the emergency event.  
We begin by presenting a 5G deployment scenario with a single densely populated industrial area, and several sparser residential areas in city area, and discuss how such a drone-based service can be implemented in the discussed area. We also present a temporal analysis of the 5G commercial traffic and emergency events based on real-life datasets~\cite{londonServices, lteTraffic}.

Due to the importance of the emergency services, reliable communications must be guaranteed. However, dedicating part of the available spectrum to the emergency services at all times can lead to severe network resource underutilization, given the sporadic nature of emergency events. Therefore, motivated by the \textit{Licensed Shared Access (LSA)}~\cite{gsmaLSA} approach according to which multiple entities have exclusive or shared access to the same spectrum in a pre-agreed manner, we propose an \textit{operator-internal shared access} scheme to reliably support emergency services on-demand, by allocating a portion of the available spectrum exclusively to the emergency services when required. 

Allocating a portion of the available bandwidth to the emergency services will impact the commercial traffic that is normally served by the base station, that heavily depends on the locality of the event, its severity, the time of the day it occurs and its duration.
For this reason, we then present a detailed spatio-temporal analysis on the impact of the proposed drone-based 5G eMBB service on commercial 5G traffic, for the first $5$ years of operation. Our analysis is based on previous work on the spatial distribution of base stations~\cite{Zhou:2015}, a publicly available commercial traffic dataset~\cite{lteTraffic}, and an extensive dataset for emergency events in the greater area of London, that spans approximately $3$ years.

The rest of the paper is organized as follows. In section~\ref{sec:deployment} we present our drone-based deployment mode for emergency services. In section~\ref{sec:analysis} we present our analysis on the spatial distribution of 5G base stations, and the temporal analysis of the commercial 5G traffic and emergency events. In section~\ref{sec:evaluation} we present our spatio-temporal correlation analysis and assess the effect of the proposed drone-based deployment on commercial 5G traffic. We conclude in section~\ref{sec:conclusions}.
\section{Deployment Model}
\label{sec:deployment}

\begin{figure}
  \centering
    \includegraphics[width=0.4\textwidth]{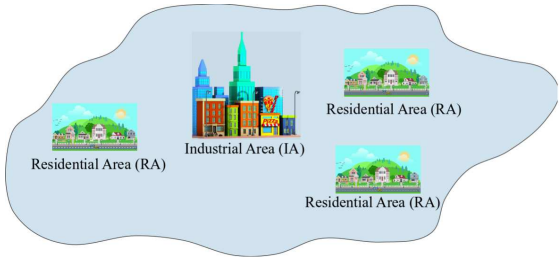}
  \caption{Example of coverage area of the city area with one industrial area (IA) and three residential areas (RAs).}
  \label{fig:cityArea}
\end{figure}

We assume a scenario where a large city area of $100km^2$ is served by multiple 4G and 5G ground cells.
Within this city area there exist a densely populated industrial area (IA), and several less densely populated residential areas (RAs) (Fig.~\ref{fig:cityArea}). We also assume that the IA can have up to $10\%$ spatial overlap with any of the RAs. However, we do not assume any area overlap among the different RAs. 
Network operators incrementally deploy 5G small cells, to cover an area of $8km^2$ every year, starting with the IA in year $1$, and adding one RA during each of the years $2$-$5$.
Each of the 5G small cells has a bandwidth of $400$ MHz, all of which is used for commercial 5G traffic in normal operating conditions. 
We also assume that a number of 4G ground small cells are being upgraded with additional 5G capabilities (e.g. additional antennas), in order to be able to provide 5G connectivity to emergency drone links.
(A cost analysis for such upgrades can be found on~\cite{costAnalPaper}.)

\begin{figure}
  \centering
    \includegraphics[width=0.5\textwidth]{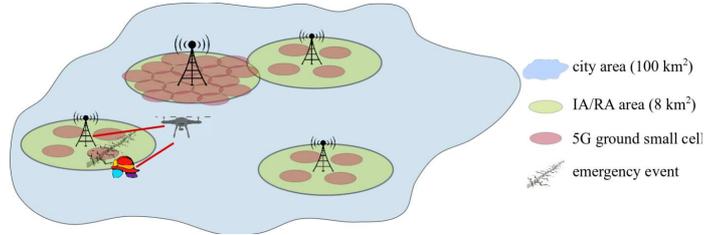}
  \caption{Drone-based scheme, where drones form a link to a ground small cell in close proximity to provide temporary connectivity to emergency services.}
  \label{fig:cityAreaEvent}
\end{figure}

We assume that localized emergency events (e.g. fires) can occur at any given time within the city area, and as such emergency communication services are required to provide connectivity to the emergency crews attending to the event. 
Towards this end, we propose to extend the limited 5G coverage (in the early stages of the deployment) with the use of drones equipped with 5G communication kits (i.e. a \textit{Remote Radio Head (RRH)}). Specifically, we assume that multiple 5G-capable drones form a connectivity link to the crews in the locality of the emergency event (Fig.~\ref{fig:cityAreaEvent}), using the 5G and/or upgraded 4G ground small cells as anchors.
Therefore, an emergency event may affect the commercial traffic only if it occurs within a 5G-covered area, as part of the spectrum of the 5G BS serving the locality of the event is allocated to the emergency services, and thus be unavailable for commercial 5G traffic. Otherwise, the drones connect to the closest upgraded ground small cell, and no effect is incurred on the commercial traffic.

Due to the importance of the emergency services, quality guarantees need to apply to the communications. 
As the drone service will be using 5G spectrum, if the event occurs within a 5G-covered area, we assume an operator-internal shared access scheme where $25\%$ of the available bandwidth (i.e. $100$ MHz) of the BS service the locality of the event is allocated to the drone link, and is unavailable for commercial traffic for the duration of the emergency event.
Although only a part of the available bandwidth is used in this case, the use of drones can provide coverage in \textit{Line-Of-Sight (LOS)} with good \textit{Signal to Noise Ratio (SNR)}. Furthermore, our analysis on a parallel paper~\cite{costAnalPaper} revealed that even when using only $25\%$ of the available spectrum, a minimum of $665$ Mbps of data rate can be achieved. As such, the sufficient data rate and reliability can be guaranteed in this case.

Based on our deployment scenario, the 5G commercial traffic is impacted only when it exceeds $75\%$ of the maximum capacity when the emergency event occurs, and therefore, the locality of the event, its occurrence time as well as duration are fundamental considerations in our analysis.
In this work we assume that city's population follows the UK core working hours, with the majority of people being in the industrial area from $9$am to $4$pm, and in the residential areas for the remaining hours of the day.

\section{Spatio-Temporal Analysis}
\label{sec:analysis}

In this section we present our spatio-temporal analysis of the proposed drone-based 5G eMBB service. We begin by describing the different datasets that we use for our analysis, and we then proceed to the spatial distribution of the base stations, and the temporal analysis of the commercial traffic and emergency events.

\subsection{Datasets Description}
\label{sec:datasets}
To develop our spatio-temporal analysis we used a commercial traffic and an emergency events dataset. 

\textbf{Commercial Traffic:} As 5G traffic datasets are not yet available, mainly due to the limited 5G deployments, for our commercial traffic analysis, we used the \textit{LTE traffic}~\cite{lteTraffic} from Google's Kaggle dataset repository, which includes downlink traffic volumes in Mbps from $57$ 4G cells, in $1$-hour intervals. The total number of entries is $9519$, spanning a whole week in October $2018$.

\textbf{Emergency Events:} For the emergency events we used detailed logs of the London fire services~\cite{londonServices}, that documented all calls to the fire services from January $2017$ to February $2019$. The dataset includes information such as the date and time of the event, the event type (e.g. major fire, flooding, false alarm), the duration of the event, and the number of pump vehicles that attended the event. In our analysis, we only considered major events (i.e. fires and floodings), and excluded specific days ($1^{st}$ January, $5^{th}$ November and $31^{st}$ December) when larger than usual number of events occur due to celebrations. Furthermore, we only considered events that required attendance of at least one pump vehicle. After clean-up, the remaining dataset consisted of $\approx41500$ entries.

\subsection{Base Station Spatial Distribution}
\label{sec:spatialDist}
As the exact BS locations in major cities are not available, several works (e.g.~\cite{Zhou:2015, Chiaraviglio:2016, Wang:2018, Li:2016}) attempt to model their distribution. In this paper, we follow the work of~\cite{Zhou:2015} and model the spatial distribution of BS for the IA and RAs using \textit{Poisson Point Process (PPP)}~\cite{ppp}. Following~\cite{Zhou:2015}, we consider different BS densities for the industrial and residential areas, using $\lambda = 53.4$ and $\lambda = 8.347$, resulting in an average number of $387$ and $61$ BSs for the industrial and residential areas respectively. 


\subsection{Traffic Temporal Distribution}
\label{sec:temporalDist}

\begin{figure}
  \centering
    \includegraphics[width=0.45\textwidth]{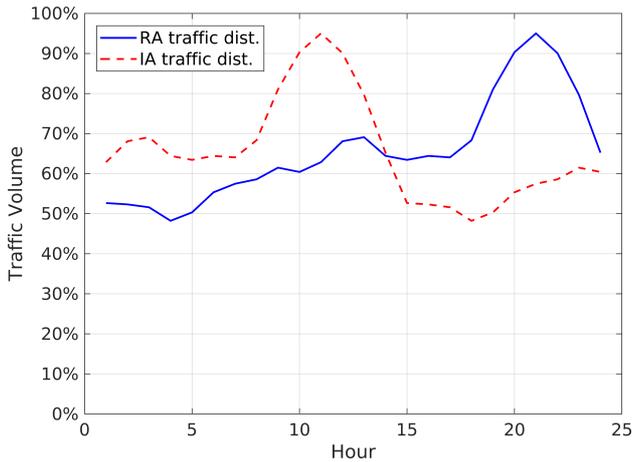}
  \caption{Temporal distribution of commercial 5G traffic for the IA and the RAs, with peaks at different core hours.}
  \label{fig:normalTrafficDist}
\end{figure}

As the \textit{LTE\_traffic}~\cite{lteTraffic} dataset spans several days, to analyze the traffic volume distribution throughout a day, we averaged the traffic volumes of each cell and each day on the respective hours. Due to the fact that the resulting traffic volumes did not reach those expected in a 5G scenario, we increased the averaged volumes to reach $95\%$ of the expected maximum capacity of a 5G small cell. The resulting traffic distribution is shown in Fig.~\ref{fig:normalTrafficDist}. Further, as the traffic volume is increased in the afternoon hours, we circularly shifted the traffic distribution by $14$ hours, to align the peak volume with the core working hours in the UK. The resulting traffic volume distribution for the industrial area is also shown in Fig.~\ref{fig:normalTrafficDist}.

\subsection{Emergency Event Temporal Distribution}
\label{sec:emergTemporalDist}
\begin{figure}
  \centering
    \includegraphics[width=0.45\textwidth]{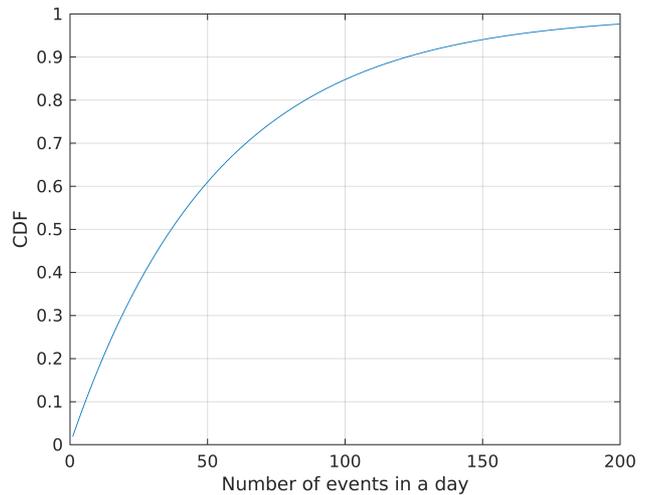}
  \caption{CDF of number of emergency events in a day.}
  \label{fig:eventsPerDay}
\end{figure}

\begin{figure}
  \centering
    \includegraphics[width=0.45\textwidth]{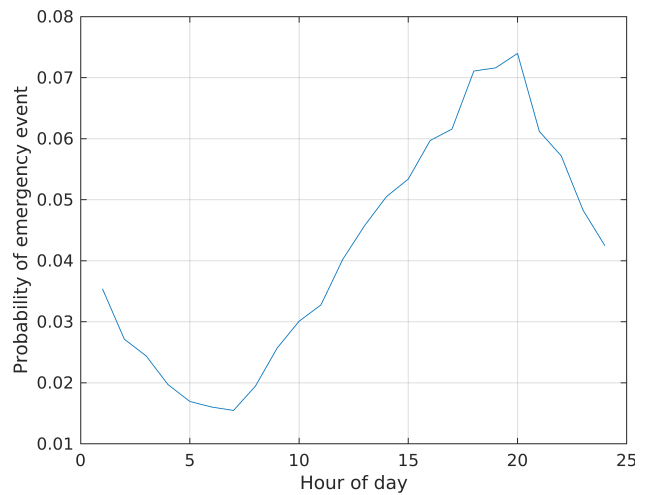}
  \caption{Probability of emergency events occurring each hour of the day. Evening hours present the higher probabilities due to the use of kitchen equipments, heating, etc.}
  \label{fig:probEmergencyOccuring}
\end{figure}

We begin by analyzing the number of events occurring each day. Fig.~\ref{fig:eventsPerDay} shows the CDF of the number of events happening within the same day, according to our emergency event dataset. The average number of emergency events in a day is $53.1$. We then analyze the probability of an event occurring at a specific time of the day (Fig.~\ref{fig:probEmergencyOccuring}). Our results show that emergency events are more likely to occur in the evening hours, when people are mainly home (e.g. cooking, heating). In fact, a close inspection on the details of the emergency events revealed that, in their majority, events were house fires.

\begin{figure}
  \centering
    \includegraphics[width=0.45\textwidth]{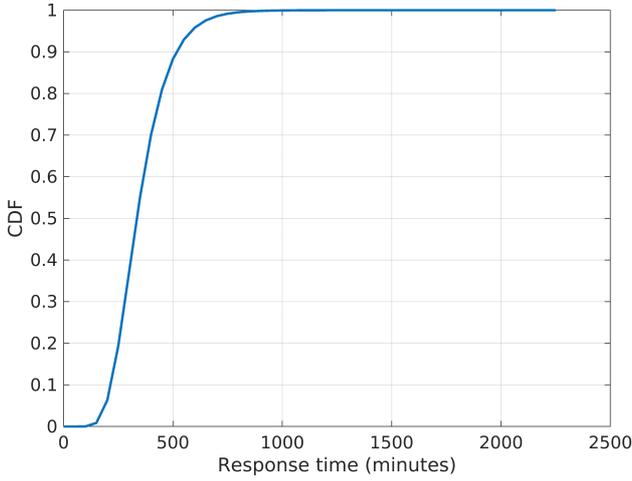}
  \caption{Temporal distribution of emergency event attending time in minutes. The maximum attending time in our dataset was $\approx19$ hours.}
  \label{fig:cdfEmergency}
\end{figure}

Finally, we analyze the duration of the emergency events (Fig.~\ref{fig:cdfEmergency}). In our dataset, the total duration of an event was indicated as the attending time (in minutes) of the pump vehicle. For the cases where more than one pump vehicles were required, we consider the total duration time of the event as the average of all the attending times, as multiple pump vehicles are operating simultaneously. The maximum attending time is $\approx19$ hours.

\section{Spatio-Temporal Correlation - Spectrum Usage Assessment}
\label{sec:evaluation}

\begin{figure}
  \centering
    \includegraphics[width=0.35\textwidth]{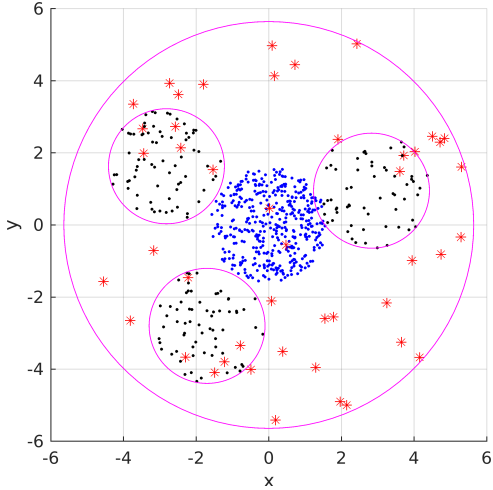}
  \caption{Example of the city area with one IA (blue dots), 3 RAs (black dots), and $53$ emergency events (red stars) randomly placed within the whole city area.}
  \label{fig:spatialDist_Events}
\end{figure}

In our scenario we assume that each day, an average number of $53$ emergency events may occur anywhere within the $100 km^2$ city area (Fig.~\ref{fig:spatialDist_Events}). The location of the emergency event (i.e. whether it occurs within a 5G-covered area or not) plays an important role in the total impact on the commercial 5G traffic (Sec.~\ref{sec:deployment}).

To assess the effect of emergency events and their support in our drone-based scenario, we define the \textit{cell impact} and the \textit{system impact}. Please note that these definitions apply only in the case where an emergency event occurs within a 5G covered area, as no impact on commercial traffic is assumed otherwise (Sec.~\ref{sec:deployment}). The cell impact determines the traffic impact on the 5G small cell that provides connectivity to the emergency services for the duration of the event (Sec.~\ref{sec:deployment}).
Further, \textit{system} we define the group of all the 5G BSs deployed an $8km^2$ area (either IA or RA), and therefore, the \textit{system impact} determines the total impact in the whole examined area. 

To determine the impact of our drone-based proposal on the commercial 5G traffic (Fig.~\ref{fig:normalTrafficDist}), we assess both the \textbf{cell} and the \textbf{system} impact given the probabilities of emergency events occurring within a 5G-covered area (IA and any of the RAs if applicable), the probability of an emergency event occurring on specific hour of the day (Fig.~\ref{fig:probEmergencyOccuring}), and the event duration (Fig.~\ref{fig:cdfEmergency}).
An important thing to note is that severe events may affect multiple small cells. 
Especially in densely populated areas (such as our considered industrial area), it is expected that 5G small cells will be deployed in significantly close proximity, or in several verticals (e.g. a multi-storey shopping center being served by a different cell in each floor)~\cite{ericsson, qualcomm}, and as such the locality of the event may extend beyond the coverage area of a single 5G small cell.
For this reason, we also analyze the severity of an events as a function of its duration. Specifically, we define $4$ severity classes, by equally splitting the maximum event duration in $4$ parts. We consider class $1$ to be the least severe, with events of that category only occurring within the coverage area of one ground small cell, while class $4$ is considered the most severe affecting $4$ ground small cells. We then derive the probabilities of an event falling into each of the different severity classes, and calculate their weighted average to determine an overall probability for the severity of an event, and the number of ground small cells it will affect.

\begin{figure}
  \centering
    \includegraphics[width=0.45\textwidth]{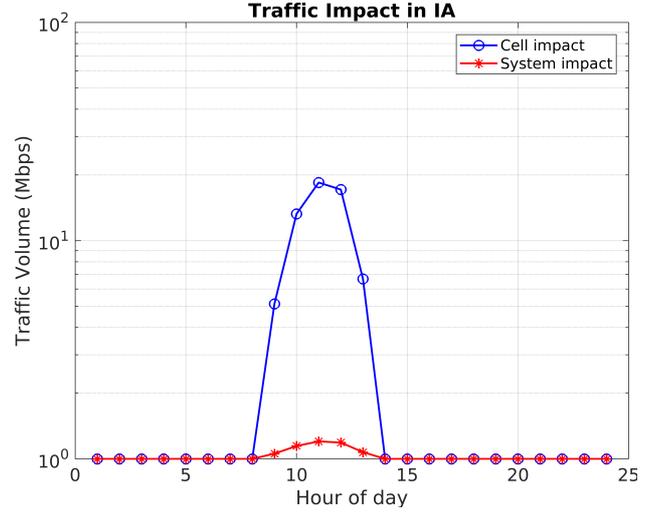}
  \caption{Cell and system impact on commercial 5G traffic in IA based on the probability of an event happening and its duration, for the industrial area.}
  \label{fig:iaImpact}
\end{figure}

\begin{figure}
  \centering
    \includegraphics[width=0.45\textwidth]{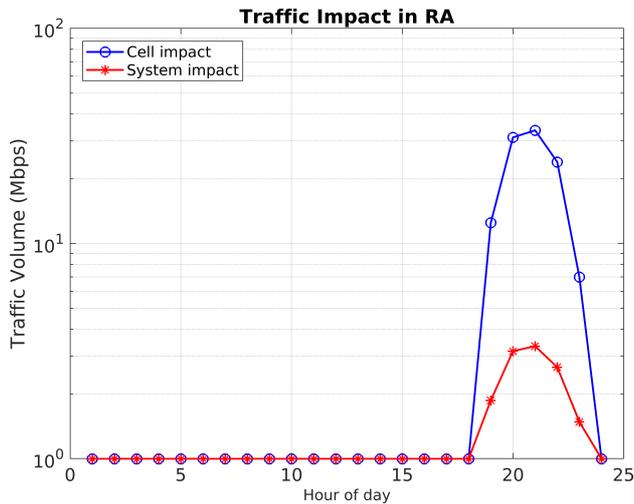}
  \caption{Cell and system impact on commercial 5G traffic in RA based on the probability of an event happening and its duration, for the industrial area.}
  \label{fig:raImpact}
\end{figure}

Based on the aforementioned assumptions and restrictions, figures~\ref{fig:iaImpact} and~\ref{fig:raImpact} depict the cell and system impact in the industrial and residential areas respectively. We can see that the cell impact in the RA is $\approx1.5$ orders of magnitude larger than that of the IA area, which is explained due to the lower probability of an event occurring there working hours. An even greater difference is also observed in the system, where the impact in the RA is $\approx2.5$ orders of magnitude larger. This is not only due to the higher probability of an emergency event occurring during the evening hours, but also due to the significantly less number of 5G smalls cells in a RA.

\begin{figure}
  \centering
    \includegraphics[width=0.42\textwidth]{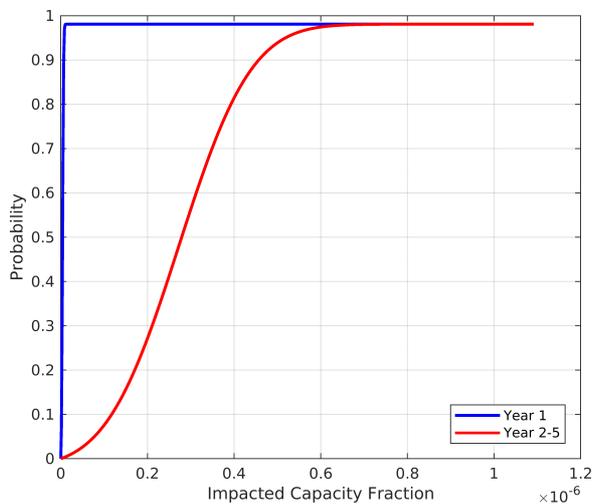}
  \caption{Total capacity impact as a fraction of the total capacity for the different years of operation.}
  \label{fig:cdfSystemImpactPerc}
\end{figure}

Finally, we assess the total capacity fraction that is impacted within all of the 5G covered areas for the first $5$ years of operation (fig~\ref{fig:cdfSystemImpactPerc}), given the average number of emergency events occurring each day. We can see that during the first year of operation, when only the IA is covered by 5G small cells, the total impact is quire small, due to the large number of 5G small cells. However, in years $2$ to $5$, 5G small cells are deployed less densely in residential areas meaning that less 5G small cells need to support the same number of emergency events, and as such the total impact increases.

\section{Conclusions \& Future Work}
\label{sec:conclusions}

In this paper we focus on the emergency services, and proposed a drone-based 5G eMBB system to provide temporary connectivity to areas with emergency events. We discuss a deployment scenario, where areas within a city are incrementally covered by 5G small cells, that are used to extend the 5G connectivity to the whole of the city area in cases of emergency by using 5G-capable drones. To provide guaranteed QoS to the emergency link, we assumed that a portion of the available spectrum is dedicated to the emergency services for the duration of the event. Then, based on our deployment scenario we presented a detailed spatio-temporal analysis on the effect of the proposed on-demand drone-based scheme on commercial 5G traffic, using real-life datasets. Our results show that the large number of 5G small cells in dense industrial areas result in a negligible impact on system capacity. In contrast, a small impact is observed in less densely populated residential areas. 
We believe that our results are encouraging towards shared access schemes even within the same operator, in 5G vertical slicing applications.
As future work, we plan to present this work to the 5GPPP spectrum working group~\cite{5gppp}, to open up further discussion on this internal LSA option.

\section*{Acknowledgements}
This work has been performed in the framework of the Horizon 2020 project ONE5G (ICT-760809) receiving funds from the European Union. The authors would like to acknowledge the contributions of their colleagues in the project, although the views expressed in this contribution are those of the authors and do not necessarily represent the project.

\small
\bibliographystyle{IEEEtran}
\bibliography{citations}

\end{document}